\documentstyle[aps,epsf,rotate,multicol]{revtex}
\begin{document}
\draft

\title{Quantifying Stock Price Response to Demand Fluctuations}

\author{Vasiliki Plerou$^{1}$, Parameswaran Gopikrishnan$^{1}$, 
Xavier Gabaix$^{2}$, and H. Eugene Stanley$^{1}$}

\address{$^{1}$ Center for Polymer Studies and Department of Physics,
Boston University, Boston, Massachusetts 02215.\\ $^{2}$ Department of
Economics, Massachusetts Institute of Technology, Cambridge,
Massachusetts 02142.\\ }

\date{\today}

\maketitle

\begin{abstract} 

We address the question of how stock prices respond to changes in
demand. We quantify the relations between price change $G$ over a time
interval $\Delta t$ and two different measures of demand fluctuations:
(a) $\Phi$, defined as the difference between the number of
buyer-initiated and seller-initiated trades, and (b) $\Omega$, defined
as the difference in number of shares traded in buyer and seller
initiated trades. We find that the conditional expectations $\langle G
\rangle_{\Omega}$ and $\langle G \rangle_{\Phi}$ of price change for a
given $\Omega$ or $\Phi$ are both concave. We find that large price
fluctuations occur when demand is very small --- a fact which is
reminiscent of large fluctuations that occur at critical points in
spin systems, where the divergent nature of the response function
leads to large fluctuations.
\end{abstract}
\pacs{PACS numbers: 05.45.Tp, 89.90.+n, 05.40.-a, 05.40.Fb}
\begin{multicols}{2}

Stock prices respond to fluctuations in demand, just as the
magnetization of an interacting spin system responds to fluctuations
in the magnetic field. Periods with large number of market
participants buying the stock imply mainly positive changes in price,
analogous to a magnetic field causing spins in a magnet to
align. Here, we quantify how price changes depend on demand
fluctuations~\cite{Zhang98,Takayasu99,Farmer99}, and find a strikingly
non-linear relationship with a specific functional form which is not
altogether unlike the dependence of magnetization on field strength.

Fluctuations in demand arise from buy and sell orders of market
participants. To quantify fluctuations in demand, we distinguish {\it
buyer-initiated} and {\it seller-initiated} trades defined by which of
the two participants in the trade, the buyer or the seller, is more
eager to trade. When such a distinction does not exist, we label the
trade as {\it indeterminate}.  We identify buyer and seller initiated
trades using the bid and ask quotes $S_{\rm B} (t)$ and $S_{\rm A}
(t)$ at which a market maker is willing to buy or sell
respectively~\cite{TAQ}. Using the mid-value $S_{\rm M} (t)= (S_{\rm
A} (t)+ S_{\rm B}(t))/2$ of the prevailing
quote~\cite{LeeReady91,Hasbrouck88,CLM}, we label a trade buyer
initiated if $S(t)> S_{\rm M}(t)$, and seller initiated if $S(t) <
S_{\rm M}(t)$. For trades occurring exactly at $S_{\rm M}(t)$, we use
the sign of the change in price from the previous trade to determine
whether the trade is buyer or seller initiated, while if the previous
trade is at the current trade price, the trade is labelled
indeterminate~\cite{LeeReady91,noteind}. Accordingly, for each trade
$i$, we define the variable
\begin{equation}
a_i \equiv \cases{ ~~1 & (buyer initiated) \cr ~~0 & (indeterminate)
\cr -1 & (seller initiated) }.
\label{defai}
\end{equation}

We quantify demand fluctuations by analyzing two quantities: (a) the
{\it number imbalance} (difference between the number of
buyer-initiated and seller-initiated trades~\cite{Jones94,Gopi00a} in
a time interval $[t, t+\Delta t]$),
\begin{mathletters}
\begin{equation}
\Phi = \Phi_{\Delta t}(t) \equiv \sum_{i=1}^{N} a_i\,,
\label{nimbalance}
\end{equation}
and (b) the {\it volume imbalance} (difference between the number of
shares traded in buyer-initiated and seller-initiated trades in a time
interval $\Delta t$),
\begin{equation}
\Omega = \Omega_{\Delta t}(t) \equiv \sum_{i=1}^{N} q_i a_i\,,
\label{vimbalance}
\end{equation}
\end{mathletters}
where $q_i$ is the number of shares traded in trade $i$, and
$N=N_{\Delta t}(t)$ is the number of trades in $\Delta t$.

\begin{figure}
\narrowtext
\centerline{
\epsfysize=0.7\columnwidth{\rotate[r]{\epsfbox{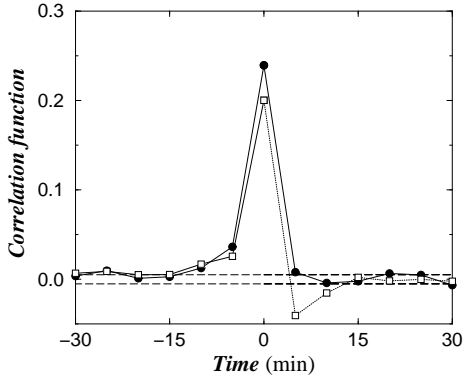}}}
}
\caption{Cross correlation functions $(\langle \Phi(t) G(t+\tau)  
\rangle - \langle \Phi(t) \rangle \langle G(t) \rangle)/\sigma_G \sigma_{\Phi}$ 
 (open circles)and $(\langle \Omega(t) G(t+\tau) \rangle - \langle \Omega(t) 
\rangle \langle G(t) \rangle)/\sigma_G  \sigma_{\Omega}$ (closed circles) computed 
using 5~min time series for $\Phi$, $\Omega$, and $G$.  We find
short-range time dependence which after $\approx 15$~min reaches noise
levels (dashed lines).}
\label{corrgqp}
\end{figure}

To choose a time scale in which to analyze the dependence of price
fluctuations on demand, we first compute the correlation functions
[Fig.~\ref{corrgqp}] $\langle \Phi(t) G(t+\tau) \rangle$ and $\langle
\Omega(t) G(t+\tau) \rangle$ and find significant dependence at $\tau=0$. 
For $\vert\tau\vert>0$, both correlation functions decay rapidly, and
cease to be statistically significant beyond $\tau \approx 15$~min ---
thereby setting a short time scale for the response of price changes
to fluctuations in demand.

\begin{figure}
\centerline{
\epsfysize=0.7\columnwidth{\rotate[r]{\epsfbox{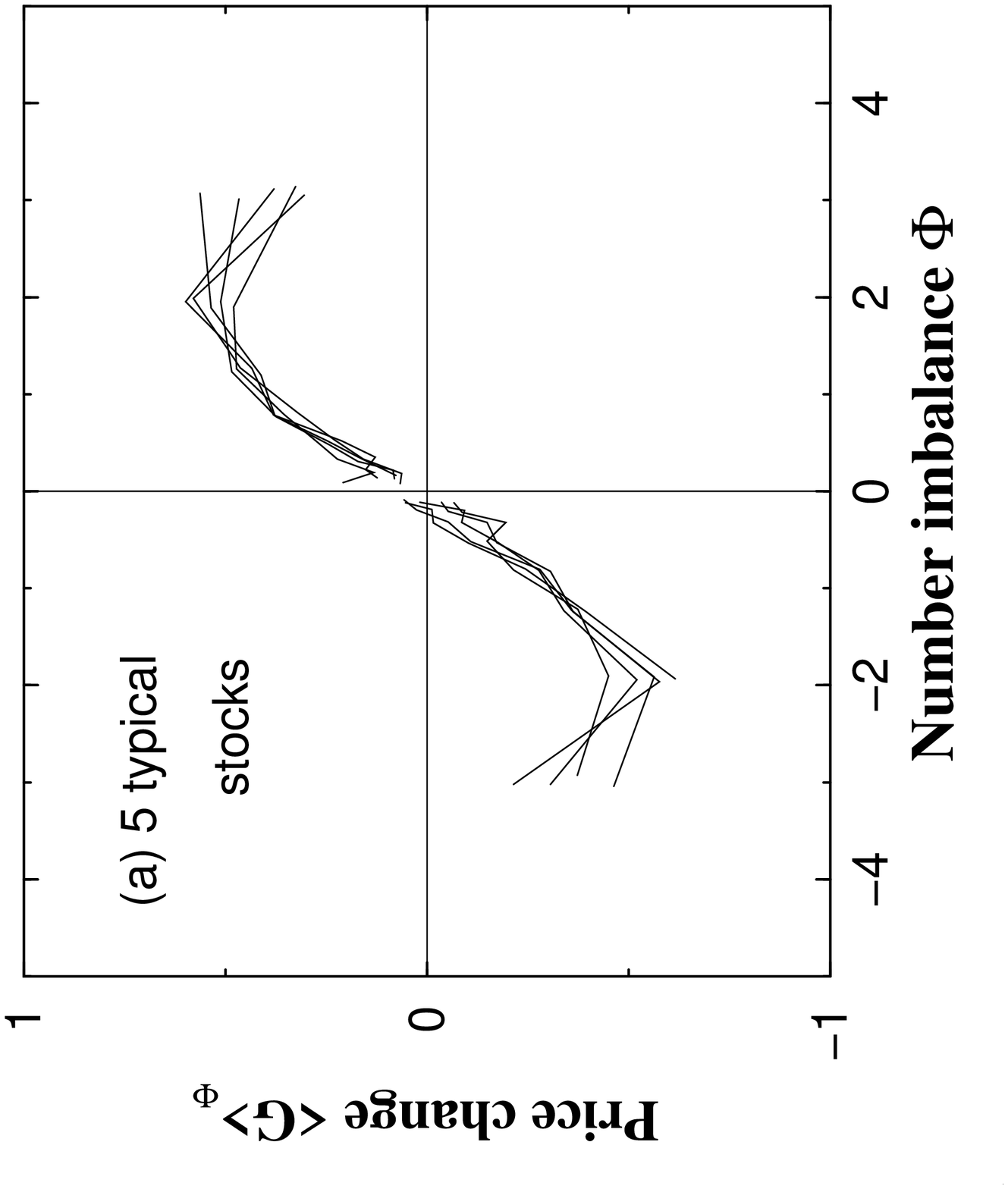}}}
}
\centerline{
\epsfysize=0.7\columnwidth{\rotate[r]{\epsfbox{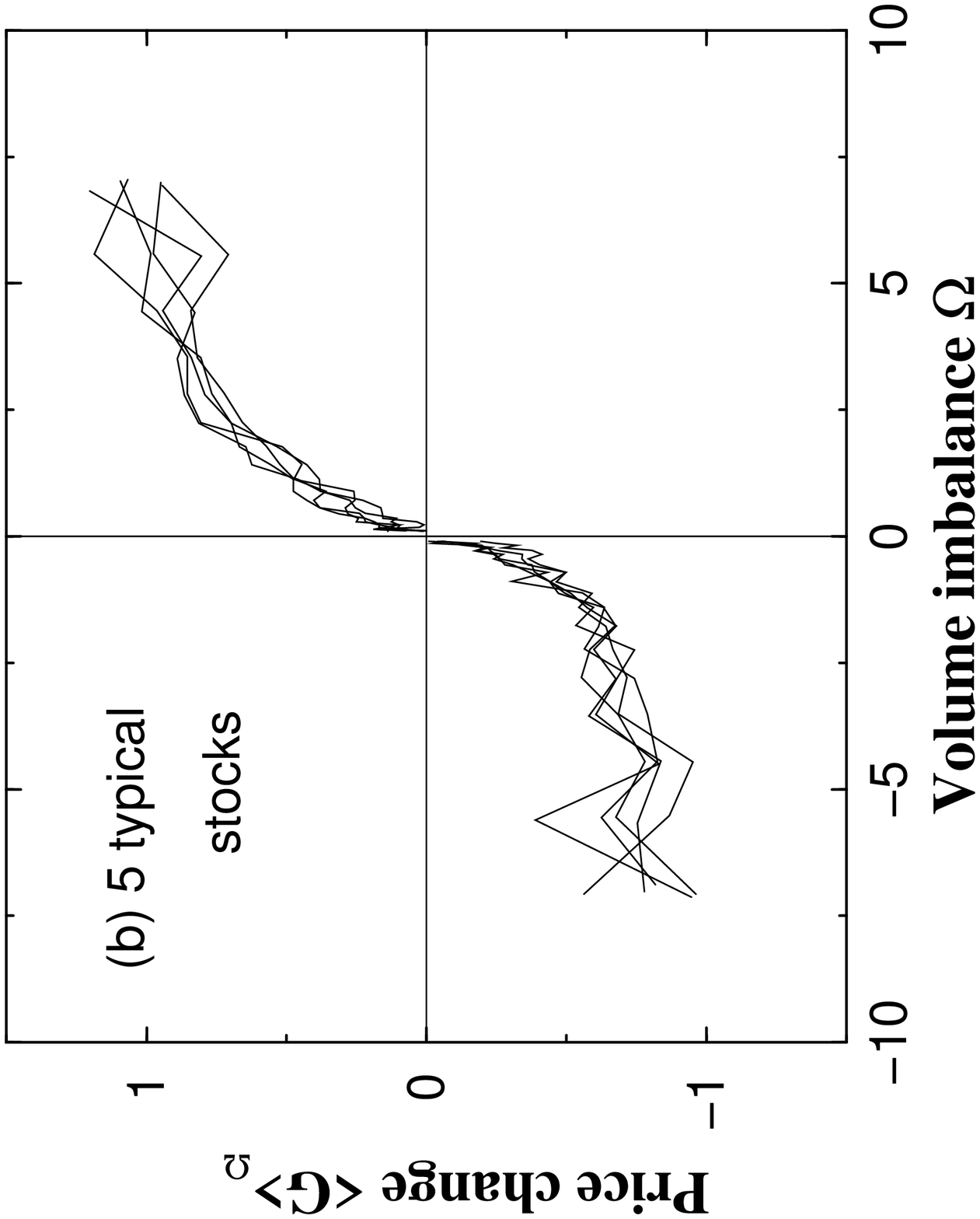}}}
}
\centerline{
\epsfysize=0.7\columnwidth{\rotate[r]{\epsfbox{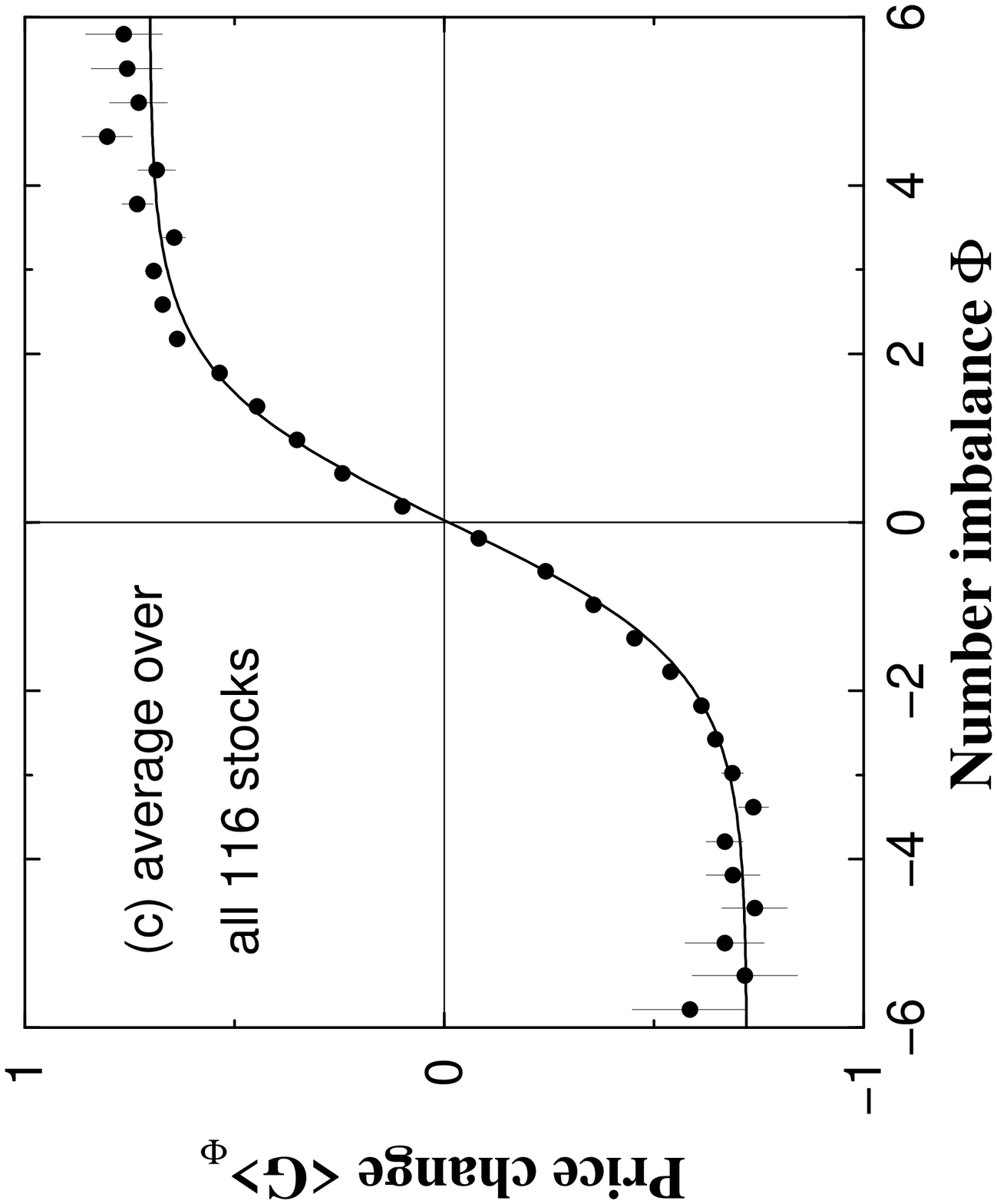}}}
}
\centerline{ 
\epsfysize=0.7\columnwidth{\rotate[r]{\epsfbox{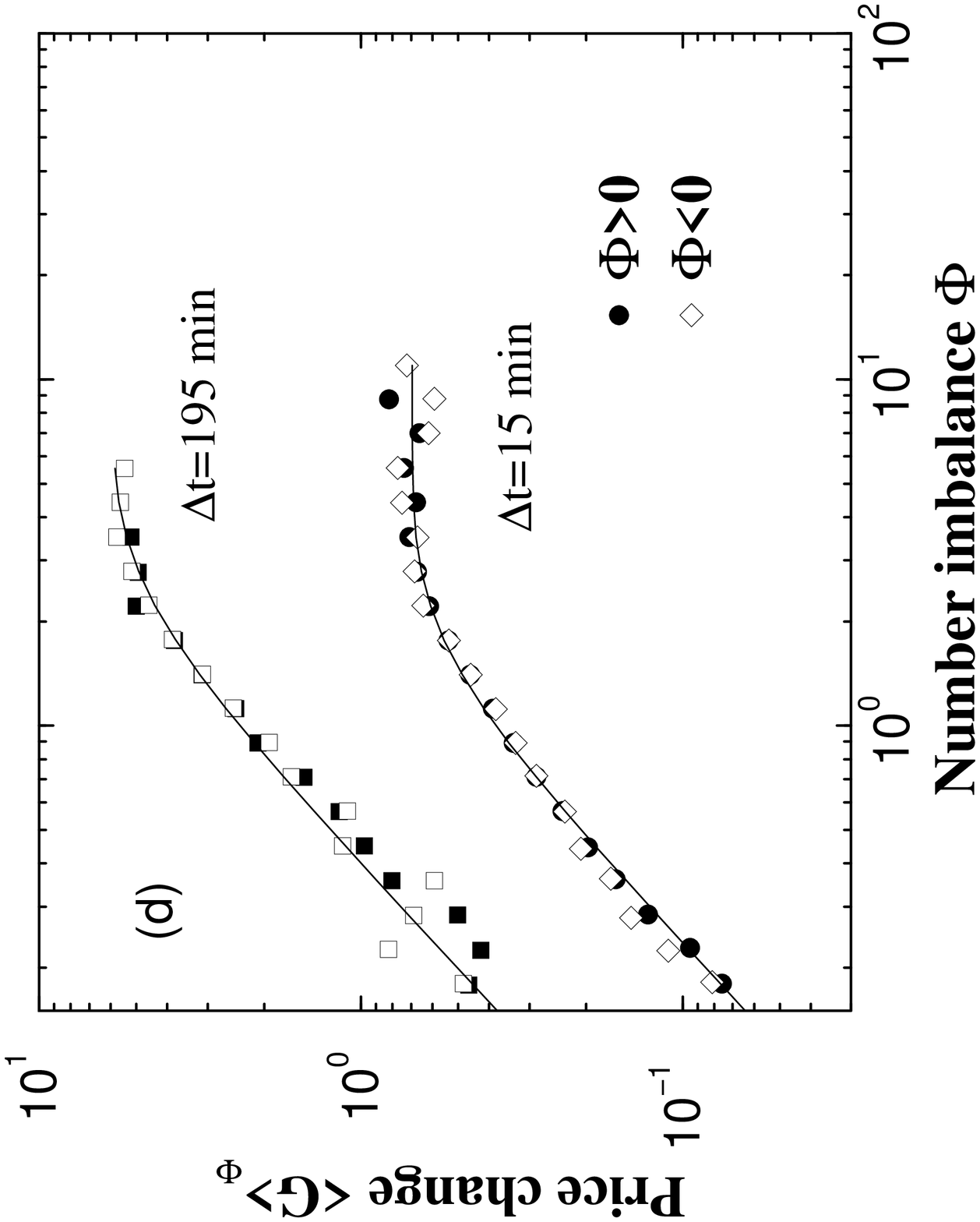}}}
}
\centerline{
\epsfysize=0.7\columnwidth{\rotate[r]{\epsfbox{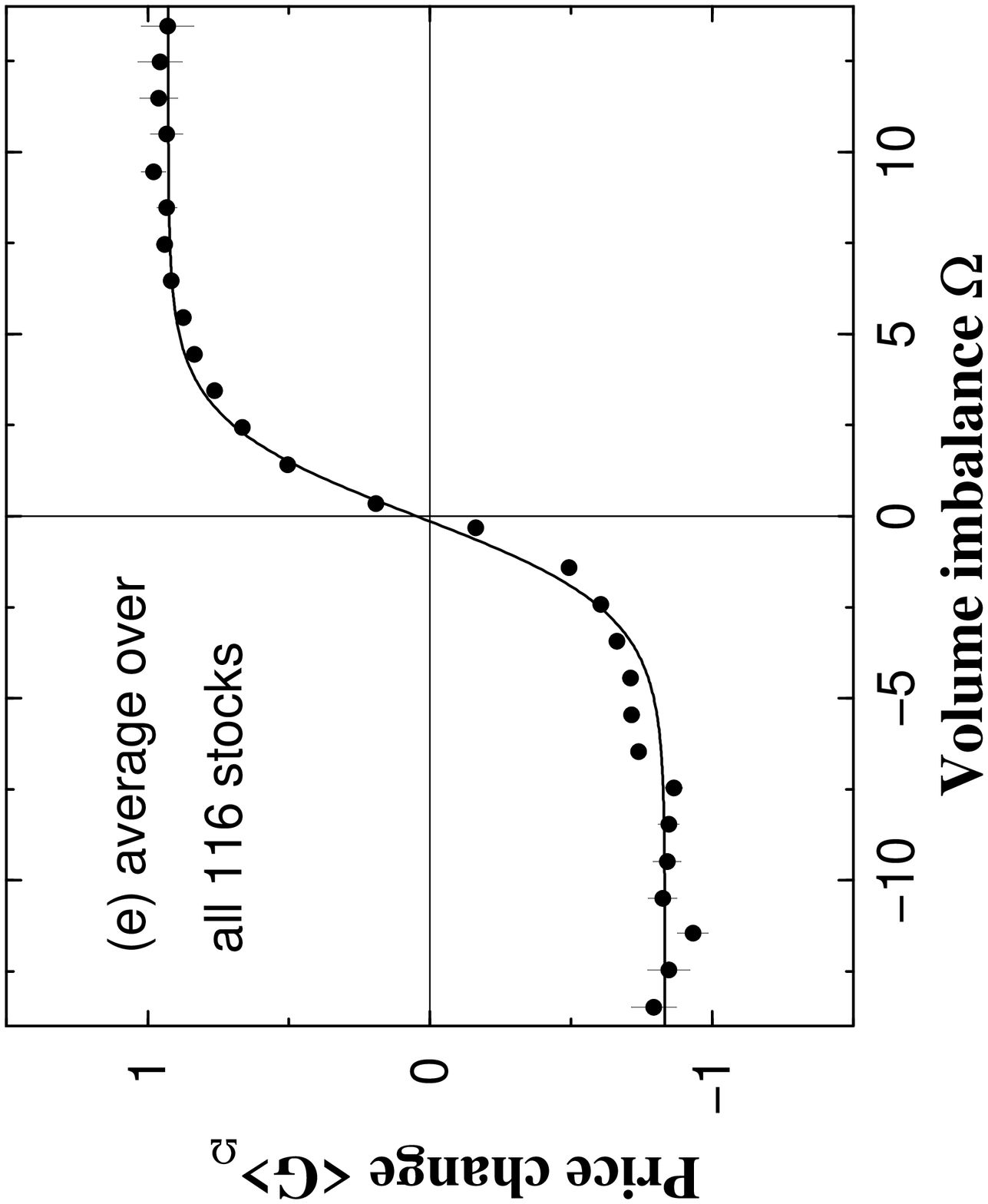}}}
}
\centerline{
\epsfysize=0.7\columnwidth{\rotate[r]{\epsfbox{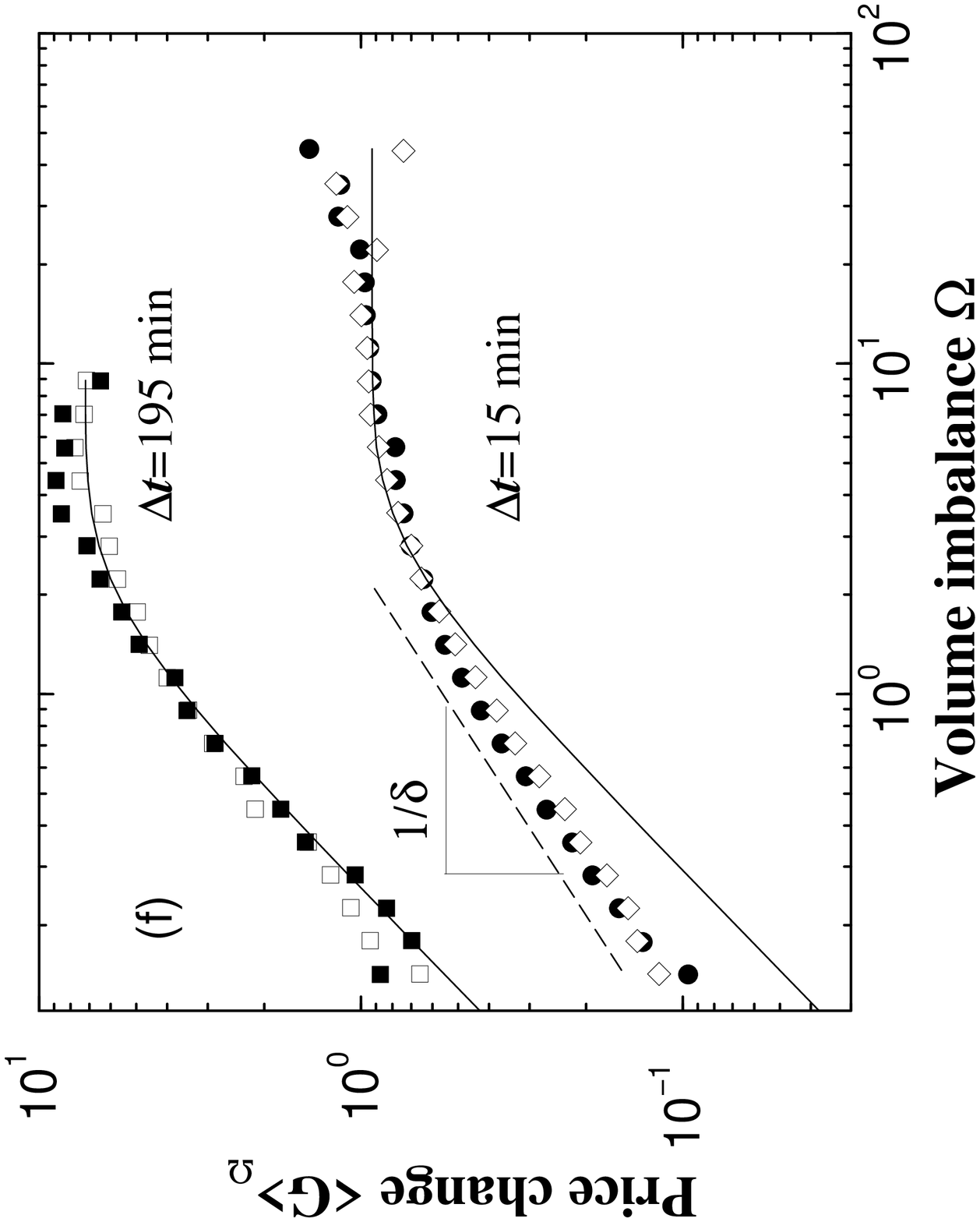}}}
}
\vspace{0.5cm}
\caption{(a) Conditional expectation $\langle G \rangle_{\Phi}$  of the price change
for a given value of $\Phi$ for 5 typical stocks over a time interval
$\Delta t=15$~min. Both $G$ and $\Phi$ are normalized to have zero
mean and unit variance. (b) Conditional expectation $\langle G
\rangle_{\Omega}$ for the same 5~stocks as in part (a).  We normalize 
$G$ to have zero mean and unit variance. Since $\Omega$ has a tail
exponent $\zeta =3/2$ which implies divergent variance, we normalize
$\Omega$ by the first moment $\langle \vert \Omega - \langle \Omega
\rangle \vert \rangle$.  (c) $\langle G \rangle_{\Phi}$ averaged over all 
116 stocks studied. The solid curve shows a fit to the function $A_0
\tanh(A_1\,\Phi)$, with $A_0=0.88 \pm 0.01$ and $A_1=0.38 \pm 0.01$,
where the fit is performed with tolerance $=0.01$. (d) Same as (c), on
a log-log plot for $\Phi>0$ (filled symbols) and $\Phi<0$ (empty
symbols) for $\Delta t=15$~min and $195$~min (shifted vertically for
clarity). The solid curves show fits to $A_0 \tanh(A_1\,\Phi)$,
which agrees well with the data.  (e) Conditional expectation $\langle
G \rangle_{\Omega}$ averaged over all 116 stocks. We calculate $G$ and 
$\Omega$ for $\Delta t=15$~min. The solid line shows a fit to the
function $B_0 \tanh (B_1 \Omega)$.  (f) $\langle G \rangle_{\Omega}$
on a log-log plot for different $\Delta t$. For small $\Omega$,
$\langle G \rangle_{\Omega} \simeq \Omega^{1/\delta}$. For $\Delta
t=15$~min find a mean value $1/\delta =0.66 \pm0.02$ by fitting
$\langle G\rangle_{\Omega}$ for all 116 stocks individually.  The same
procedure yields $1/\delta = 0.34 \pm 0.03$ at $\Delta t=5$~min
(interestingly close to the value of the analogous critical exponent
in mean field theory). The solid curve shows a fit to the function
$B_0\,\tanh(B_1\,\Omega)$. For small $\Omega$,
$B_0\,\tanh(B_1\,\Omega) \sim \Omega$, and therefore disagrees with
$\langle G \rangle_{\Omega}$, whereas for large $\Omega$ the fit shows
good agreement. For $\Delta t=195$~min (${1\over 2}$~day) (squares),
the hyperbolic tangent function shows good agreement. }
\label{gcondqp}
\end{figure}

Next, we shall examine the relationships
\begin{mathletters}
\begin{equation}
\langle G \rangle_{\Phi} \equiv {\bf E} (G\vert \Phi)\,,
\label{defF}
\end{equation}
\begin{equation}
\langle G \rangle_{\Omega} \equiv {\bf E} (G\vert \Omega)\,,
\label{defF1}
\end{equation}
\end{mathletters}
which give the equal-time expectation value of price change $G(t)$ for
a given $\Phi(t)$ or $\Omega(t)$. Figures~\ref{gcondqp}(a) and (b)
show $\langle G \rangle_{\Phi}$ and $\langle G \rangle_{\Omega}$ for 5
typical stocks for $\Delta t=15$~min. We find that both $\langle G
\rangle_{\Phi}$ and $\langle G \rangle_{\Omega}$ are nonlinear, displaying
concave curvature with increasing $\Phi$ and
$\Omega$~\cite{Chan93,Keim94,Haussman91}, and `flattening' at large
values~\cite{notemidq}.

Figure~\ref{gcondqp}(c) shows the average behavior of $\langle G
\rangle_{\Phi}$ for all stocks.  We find that $\langle G \rangle_{\Phi}$ 
is consistent with the functional form
\begin{equation}
\langle G \rangle_{\Phi} = A_0 \,\tanh (A_1 \Phi)\,,
\label{tanh}
\end{equation}
where $A_0$ is a constant that denotes the level of `saturation', and
$A_1$ determines the average price change for unit change in $\Phi$.
In the case of a spin system, the saturation at large values for the
analogous curve --- magnetization vs. field --- is due to the fixed
number of spins. The apparent saturation of $\langle G
\rangle_{\Phi}$ is surprising in the present
context, since there is no clear upper limit either on the price
change, or on the number of trades. We find that $\langle G
\rangle_{\Phi}$ for a range of $\Delta t$, also displays good 
agreement with Eq.~(\ref{tanh}) [Fig.~\ref{gcondqp}(d)].

We next focus on $\langle G \rangle_{\Omega}$
[Fig.~\ref{gcondqp}(e)]. We find that the function $\langle G
\rangle_{\Omega}$, like $\langle G \rangle_{\Phi}$, is consistent with
Eq.~(\ref{tanh})~\cite{Zhang98,notebound,notegqp}. However, near
$\Omega=0$, $\langle G \rangle_{\Omega}$ shows not a strict linear
behavior for small $\Omega$ as we expect for $\tanh \Omega$, but
rather a power-law $\langle G \rangle_{\Omega} \sim
\Omega^{1/\delta}$ [Fig.~\ref{gcondqp}(e)]. We find that $1/\delta$
depends on $\Delta t$ [Fig.~\ref{gcondqp}(f)]: $\delta \approx 3$ for
$\Delta t=5$~min and $\delta \approx 3/2$ for $15$~min, and $\delta
\rightarrow 1$ for larger $\Delta t$, agreeing well with $\tanh
\Omega$ ~\cite{notebound}.

\begin{figure}
\narrowtext
\centerline{
\epsfysize=0.7\columnwidth{\rotate[r]{\epsfbox{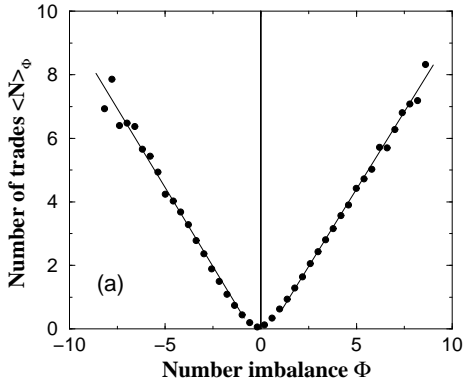}}}
}
\centerline{
\epsfysize=0.7\columnwidth{\rotate[r]{\epsfbox{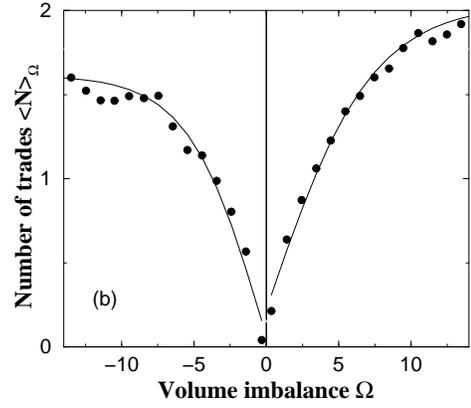}}}
}
\caption{(a) Conditional expectation  $\langle N \rangle_{\Phi}$  
of the number of trades for a given $\Phi$ averaged over all 116
stocks, shows approximately {\underline linear} behavior with
increasing $\Phi$. (b) $\langle N \rangle_{\Omega}$ averaged over all
116 stocks shows strikingly {\underline nonlinear} behavior. The solid
line shows a fit to the function $C_0 -C_1 \exp (-C_2 \Omega)$ (which
has the same large $\Omega$ behavior as a hyperbolic tangent). For both
parts (a) and (b) We calculate $G$, $\Phi$ and $\Omega$ over $\Delta
t=15$~min.  Both $\Phi$ and $G$ are transformed to have zero mean and
unit variance, whereas $\Omega$ is normalized by its first moment.}
\label{nCqp}
\end{figure}

Next, we analyze the dependence of the number of trades $N$ on demand
fluctuations to quantify how large volume imbalances generate
trades. Figure~\ref{nCqp}(a) shows that the equal-time expectation
value $\langle N \rangle_{\Phi}$ shows a linear increase with
$\Phi$. The dependence of $N$ on volume imbalance $\Omega$ is
nonlinear; $\langle N \rangle_{\Omega}$ displays a ``cusp'' at
$\Omega=0$ followed by a sharp increase and saturation at large values
[Fig.~\ref{nCqp}(b)].

\begin{figure}
\narrowtext
\centerline{
\epsfysize=0.7\columnwidth{\rotate[r]{\epsfbox{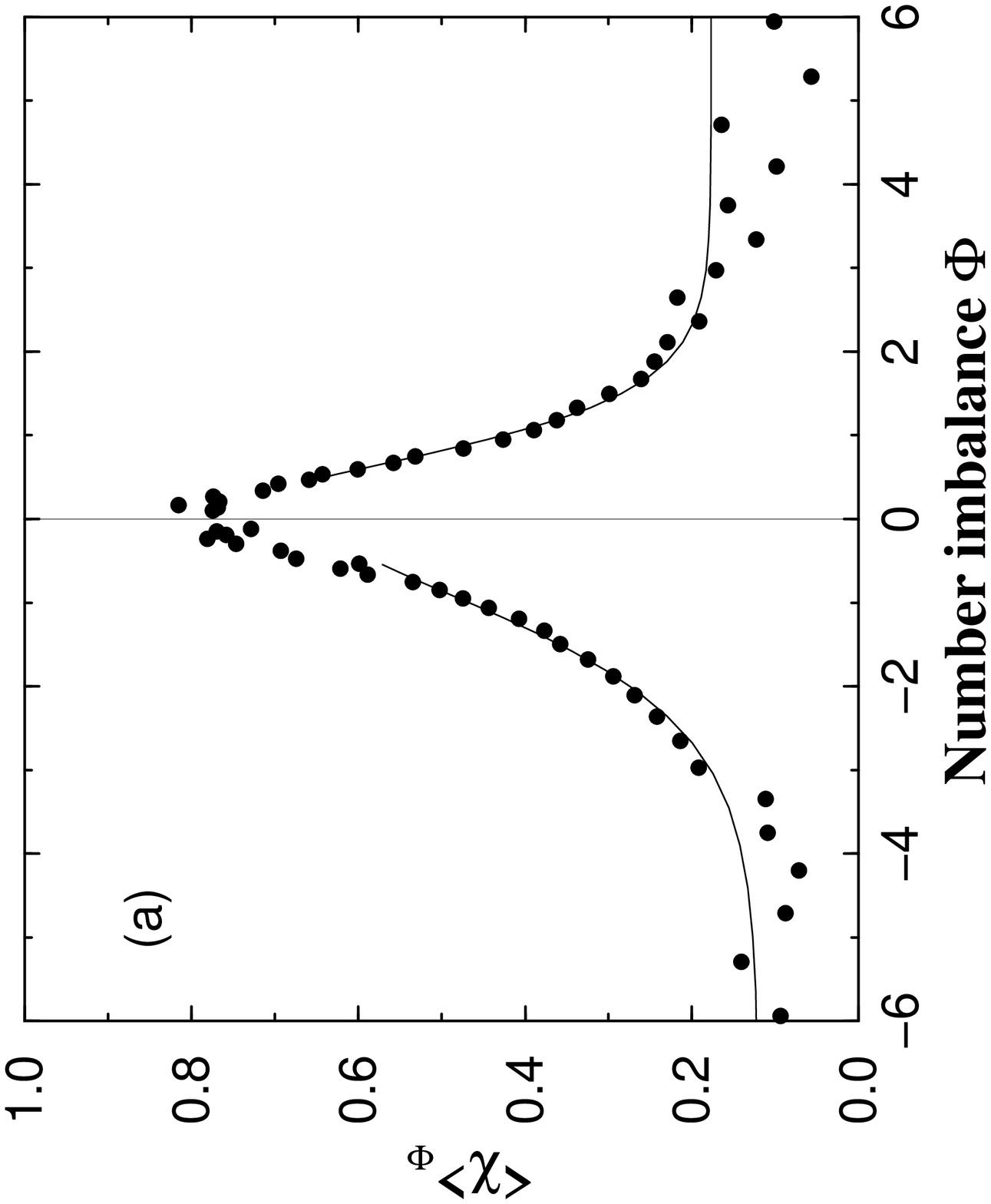}}}
}
\centerline{
\epsfysize=0.7\columnwidth{\rotate[r]{\epsfbox{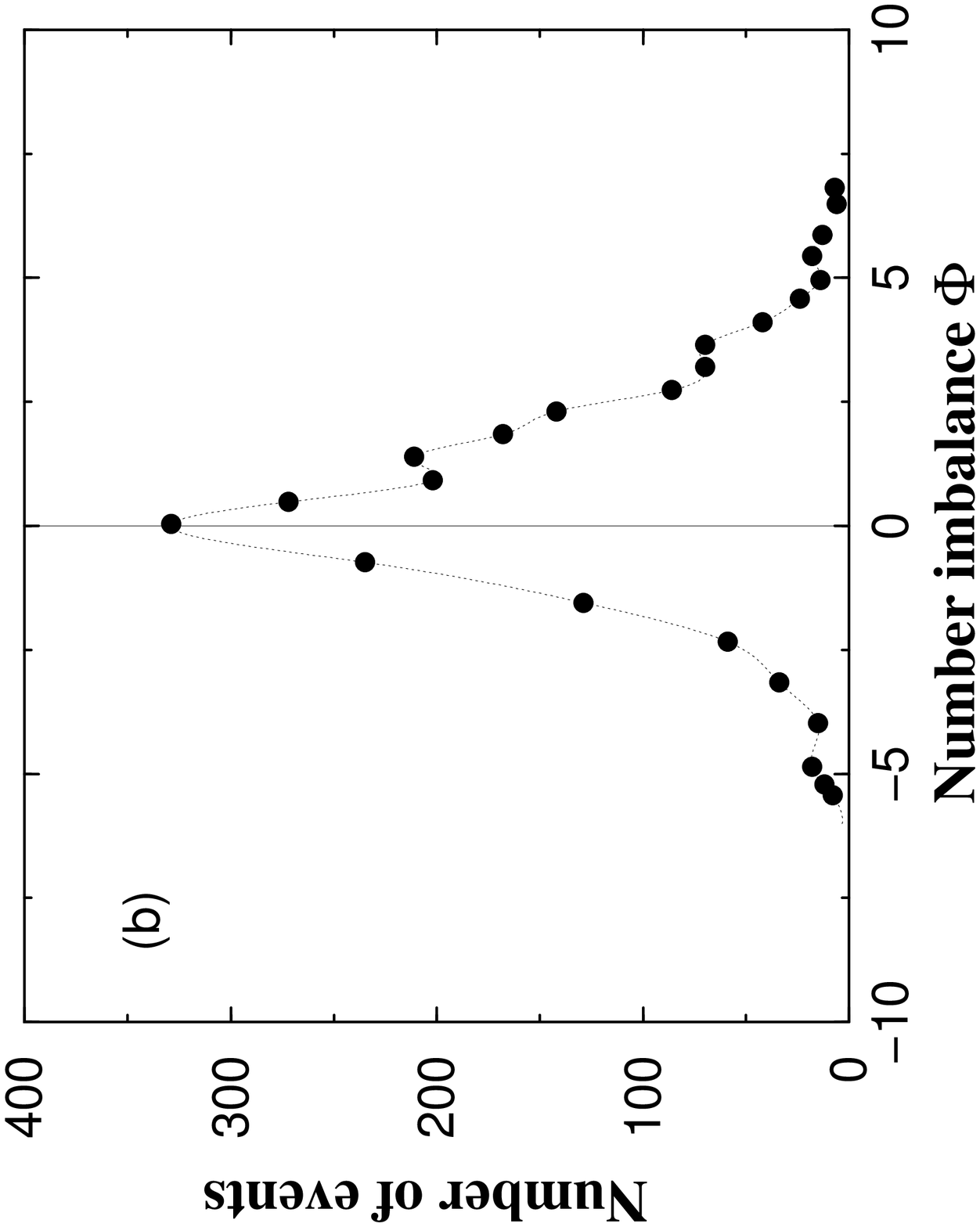}}}
}
\caption{(a) Conditional expectation $\langle \chi \rangle_{\Phi}$,
where $\chi$ is calculated using Eq.~(\protect\ref{defchi}), shows
large values near $\Phi=0$ and decay for increasing $\Phi$. The
solid lines show a fit to the function $D_0 {\rm sech}^2(D_1
\Phi)$. (b) Number of events with $\vert G \vert >5$~standard 
deviations for a given $\Phi$ shows large values at $\Phi=0$.}
\label{chi}
\end{figure}

In spin systems, the amplitude of spin fluctuations is related to the
susceptibility, which quantifies the response of the system to
fluctuations in the magnetic field. In our problem, a certain change
$\Delta \Phi$ in demand $\Phi$ (analog of the field) causes a response
${\delta \langle G \rangle_{\Phi} \over \delta \Phi} \Delta
\Phi$, which we find to be largest at $\Phi=0$ [Fig.~\ref{gcondqp}],
suggesting that the non-linear shape of $\langle G \rangle_{\Phi}$ can
give rise to large fluctuations (large ``volatility''~\cite{Liu97})
when $\Phi$ is small. The average amplitude of fluctuations in
$G\equiv\sum_{i=1}^N \delta p_i$ is given by the variance
\begin{equation}
\chi^2 \equiv \langle\delta p_i^2\rangle - \langle \delta p_i \rangle^2\,,
\label{defchi}
\end{equation}
where $\delta p_i$ is the price change due to trade $i$ and
$\langle\dots \rangle$ denotes the average computed over the interval
$\Delta t$. Figure~\ref{chi}(a) shows that $\langle \chi
\rangle_{\Phi}$ displays large values near $\Phi=0$ and a rapid decay
for increasing $\Phi$. Figure~\ref{chi}(b) shows as a function of
$\Phi$ the number of events with price change $\vert G
\vert>5$~standard deviations. Interestingly, we find that a majority of 
the large events occur at $\Phi=0$, consistent with previous empirical
results~\cite{Plerou00} which show that the power-law distribution of
price changes~\cite{Lux} mainly arises from $\chi$.  Our findings are
reminiscent of phase transitions in spin systems, where the divergent
behavior of the response function at the critical point (zero magnetic
field) leads to large fluctuations~\cite{Stanley71}.

We thank L.~A.~N.~Amaral, J-P.~Bouchaud, K.~B.~Doran, M.~Meyer, and
B.~Rosenow for helpful conversations, and the National Science
Foundation for financial support.


\vspace{-3cm}

\end{multicols}


\begin{references}
\bibitem{notemidq}
To test that short-term anticorrelation in price changes (``bid-ask
bounce'') do not affect the conditional expectation values, we have
also used mid-quote changes and find similar results.
\bibitem{notebound}
Regardless of the fitting function, the `flat' asymptotic behavior of
$\langle G \rangle_{\Omega}$ is consistent with what we expect from the
distributions of $G$ and $\Omega$. Since $q_i$ has a distribution
$P_q\{q >x\} \sim x^{-\zeta}$ with exponent $\zeta\approx 3/2$ within
the L\'evy stable domain~\cite{Gopi00a}, the distribution $P_{\Omega}
\{\Omega>x\}$ also has the same tail exponent, since the variables
$q_i$ and $a_i$ in Eq.~(\ref{vimbalance}) have only short-ranged time
dependence. Price fluctuations have distribution $P_G \{G>x\} \sim
x^{-\alpha}$ with $\alpha \approx 3$~\cite{Lux}. Therefore, the
asymptotic behavior of the function $\langle G \rangle_{\Omega}$ must
be bounded from above by $\Omega^{1/2}$, since for $\langle G
\rangle_{\Omega} \sim \Omega^{\delta}$ with $\delta>1/2$, the 
exponent $\alpha$ of $P_G(G>x)$ must be smaller than the empirical
value of $\alpha\approx 3$. Thus, for consistency of tail exponents of
$G$ and $\Omega$, we must have $\delta \leq 1/2$ for large $\Omega$,
which is consistent with the `flat' behavior of $\langle G
\rangle_{\Omega}$ for large $\Omega$ that we
find. Zhang~\cite{Zhang98} presents an argument for $\langle G
\rangle_{\Omega} \sim \Omega^{1/2}$.
\bibitem{notegqp} Although we fit $\langle G \rangle_{\Omega}$ with 
$\tanh (\Omega)$, there are alternative functions that fit $\langle G
\rangle_{\Omega}$. From Fig.~\ref{gcondqp}(e), it is evident
that $\langle G \rangle_{\Omega}$ is weaker than any power of
$\Omega$. In particular, we also obtain good fits using the function
$\langle G \rangle_{\Omega} = A_0\log (A_1+A_2\vert \Omega \vert)$.
\end{references}
\end{document}